\documentclass[aps,prd,nofootinbib,twocolumn,reprint,groupedaddress,showpacs,showkeys]{revtex4-1}

\usepackage{graphicx}  
\usepackage{dcolumn}   
\usepackage{bm}        
\usepackage{amssymb}   
\usepackage{color}
\usepackage{multirow}
\usepackage {amsmath}
\usepackage {amsfonts}
\usepackage {amsthm} 
\usepackage {mathrsfs}
\usepackage {natbib}
\usepackage {latexsym}
\usepackage {dsfont}
\usepackage {txfonts}
\usepackage {rotating}
\usepackage {wasysym}
\usepackage {multirow}
\usepackage {hhline}
\usepackage {hyperref}
\usepackage {bm}
\usepackage {appendix}
\usepackage {url}
\usepackage {acronym}
\usepackage{float}

\newcommand{\tn}{\textnormal}

\begin{document}

\title{Low latency search for Gravitational waves from BH-NS binaries
in coincidence with  Short Gamma Ray Bursts}

\author{Andrea Maselli}
\affiliation{Dipartimento di Fisica, Universit\`a di Roma ``La Sapienza'' \& Sezione, INFN Roma1, P.A. Moro 5, 00185, Roma, Italy.}
\email{andrea.maselli@roma1.infn.it}

\author{Valeria Ferrari}
\affiliation{Dipartimento di Fisica, Universit\`a di Roma ``La Sapienza'' \& Sezione, INFN Roma1, P.A. Moro 5, 00185, Roma, Italy.}
\email{valeria.ferrari@roma1.infn.it}

\date{\today}

\begin{abstract}
We propose a procedure to  be used in the search for
gravitational waves from black hole-neutron star coalescing
binaries, in coincidence with short gamma-ray bursts.
It is based on two recently proposed semi-analytic fits, 
one reproducing the mass of the remnant disk surrounding the black hole
which forms after the merging as a function of some binary parameters,
the second relating the neutron star compactness, i.e. the
ratio of mass and radius, with its tidal deformability. Using a Fisher
matrix analysis and the two fits, we
assign a probability that the emitted  gravitational signal
is associated to the
formation of an accreting disk massive enough to supply the energy
needed to power a short gamma ray burst. This information  
can be used in low-latency data analysis
to restrict the parameter space searching for gravitational wave signals
in coincidence with short gamma-ray bursts, 
and to gain information on the dynamics of
the coalescing system and on the internal structure of
the components. In addition, when the binary parameters will be measured
with high accuracy, it will be possible to use this information  to
trigger the search for off-axis gamma-ray bursts afterglows.

\end{abstract}


\maketitle

{\em Introduction.}---
The advanced gravitational wave detectors  LIGO
and Virgo (to hereafter AdvLIGO/Virgo) are expected to detect signals
emitted by coalescencing compact binaries, formed by neutron stars (NS)
and/or black holes (BH) \cite{LIGOVirgo}.  These catastrophic events
have an electromagnetic counterpart. For instance,
the coalescence of NS-NS and BH-NS binaries has been proposed
as a candidate for the central engine of short Gamma Ray Bursts (SGRB),
provided the stellar-mass BH which forms after merging is surrounded by
a hot and sufficiently massive accreting disk, but this model needs to
be validated (see for instance \cite{LR07} and references therein).  
Since the electromagnetic emission is produced at large
distance from the central engine, it does not give strong information on
the source. In addition, the emission is beamed, and consequently these
events may not be detected if one is looking in the wrong direction.
Conversely, the gravitational wave (GW) emission is not beamed, and
exhibits a characteristic waveform (the chirp) which should allow a non
ambiguous
identification of the source. GRBs are characterized by a prompt 
emission, which lasts a few seconds, and an afterglow, whose duration 
ranges from hours to days. 

Thus, gravitational wave detection may be
used to trigger the afterglow search of  GRBs
which have not been detected by the on-axis prompt observation, and to
validate  the ``jet-model'' of SGRB.
Or, in alternative, the observation of a SGRB may be used as a trigger
to search for a coincident GW signal. Indeed, this kind of search has
already been done in the data of LIGO and Virgo \cite{Abadieb,Abadiec}.

Since not all coalescences of compact bodies produce a black hole 
with an accreting disk sufficiently massive to power a SGRB,
we need to devise  a strategy to extract
those having the largest probability 
to produce a SGRB. This is one of the purposes of this paper.
In a recent paper of the LIGO-Virgo collaboration \cite{LIGOVirgoplan} a 
plausible observing schedule has been indicated, according to which within 
this decade the advanced detectors, operating under appropriate conditions, 
will be able to determine the sky location of a source within $5$ and $20$ 
deg$^2$. Given the cost of spanning this quite large region of sky  to search 
for a coincident SGRB with electromagnetic detectors, indications on whether 
a detected signal is likely to be associated 
with a SGRB is a valuable information. 

The procedure we propose has several applications. It can be
used in the data analysis of future detectors  i)
to gain information on the range of parameters which is more useful to
span in the low latency search for GWs emitted by BH-NS sources
\cite{Abadiea},  ii) for an externally triggered search for GW
coalescence signals, following GRB observations \cite{Abadieb,Abadiec},
and iii) when the binary parameters will be measured with sufficient
accuracy and in a sufficiently short time to allow for an
electromagnetic follow-up, to search for off-axis GRB afterglows.
Although our  method is devised for BH-NS coalescing binaries, it  will
also be applicable to NS-NS binaries, when a reliable and suitable fit
for the mass of the accretion disk which forms around the black hole
produced in the coalescence will be provided by numerical studies of
such systems (see below).

A large number of numerical studies of BH-NS coalescence, 
have allowed to derive two interesting fits. The first 
\cite{F12} gives the mass of the accretion disk, $M_{\tn{rem}}$, 
as a function of the the NS compactness
${\cal C}=M_{\tn{NS}}/R_\tn{NS}$, where $M_{\tn{NS}}$ and $R_\tn{NS}$
are the NS mass and radius,
the dimensionless BH spin, $\chi_{\tn{BH}}\in[-1,1]$, and the mass ratio
$q=M_{\tn{BH}}/M_{\tn{NS}}$:
\begin{equation}\label{torusFit}
\frac{M_{\tn{rem}}}{M_{\tn{NS}}^{b}}=K_1 (3q)^{1/3}(1-2{\cal C})
-K_2 q\ {\cal C}\ R_{\tn{ISCO}}\ .
\end{equation}
Here $M^\tn{b}_\tn{NS}$  is the NS baryonic mass which,
following \cite{GPRTL13}, we assume to be  10\% larger than the 
NS gravitational mass;
$R_{\tn{ISCO}}$ is the radius of the innermost, stable circular orbit 
for a Kerr black hole:
\begin{equation}
\frac{R_{\tn{ISCO}}}{M_{\tn{BH}}}=3+Z_{2}-\tn{sign}(\chi_{\tn{BH}})
\sqrt{(3-Z_{1})(3+Z_{1}+2Z_{2})}\ ,
\end{equation}
where 
$ Z_{1}=1+(1-\chi_{\tn{BH}}^{2})^{1/3}\left[(1+\chi_{\tn{BH}})^{1/3}
+(1-\chi_{\tn{BH}})^{1/3}\right]$ and $
Z_{2}=(3\chi_{\tn{BH}}+Z_{1}^{2})^{1/2}$
\cite{BPT72}.
The two coefficients $K_1=0.288\pm0.011$ and $K_2=0.1248\pm 0.007$ have
been derived \cite{F12} through a least-square fit of the results of fully
relativistic numerical simulations 
\cite{KOST11,ELSB09,FDKSST12,FDKT11}.

$M_{\tn {rem}}$ is a key parameter in our study.
Neutrino-antineutrino annihilation processes extract energy from the
disk \cite{Piran04}, and several studies have shown that
this process could supply the energy required to ignite
a short gamma-ray burst, if 
$M_{\tn {rem}}\in(0.01\div0.03)M_{\odot}$ \cite{SRJ04,MPN02,PWF98}. 
In the following we
shall assume as a threshold for SGRB formation $M_{\tn
{rem}}=0.01~M_{\odot}$. The results we will show do not change if we
choose  $M_{\tn {rem}}=0.03~M_{\odot}$. 
 
The second fit \cite{MCFGP13} is a universal relation between the NS compactness
${\cal C}$ and the tidal deformability
$ \lambda_{2}=-Q_{ij}/C_{ij}$, where
$Q_{ij}$ is the star traceless quadrupole tensor, and
$C_{ij}=e^\alpha_{(0)}e^\beta_{(i)}e^\gamma_{(0)}e^\delta_{(j)}R_{\alpha\beta\gamma\delta}$
is the tidal tensor, i.e. the Riemann
tensor projected onto 
the parallel transported tetrad attached to the star
$e^\alpha_{(\mu)}$:
\begin{equation}\label{Clambda} 
{\cal C}=0.371-3.9\times 10^{-2}\ln \bar{\lambda}
+1.056\times 10^{-3}(\ln\bar{\lambda})^2\ ,
\end{equation}
where $\bar{\lambda}=\lambda_{2}/M_{\tn{NS}}^5$.
This fit is found to reproduce the values of the star
compactness with
accuracy greater 3\%, for a large class of equations of state (EoS). 
Hereafter, we shall
denote by
${\cal C}_\lambda$ the NS compactness obtained from this fit.

\begin{figure}[ht]
\includegraphics[width=6.5cm]{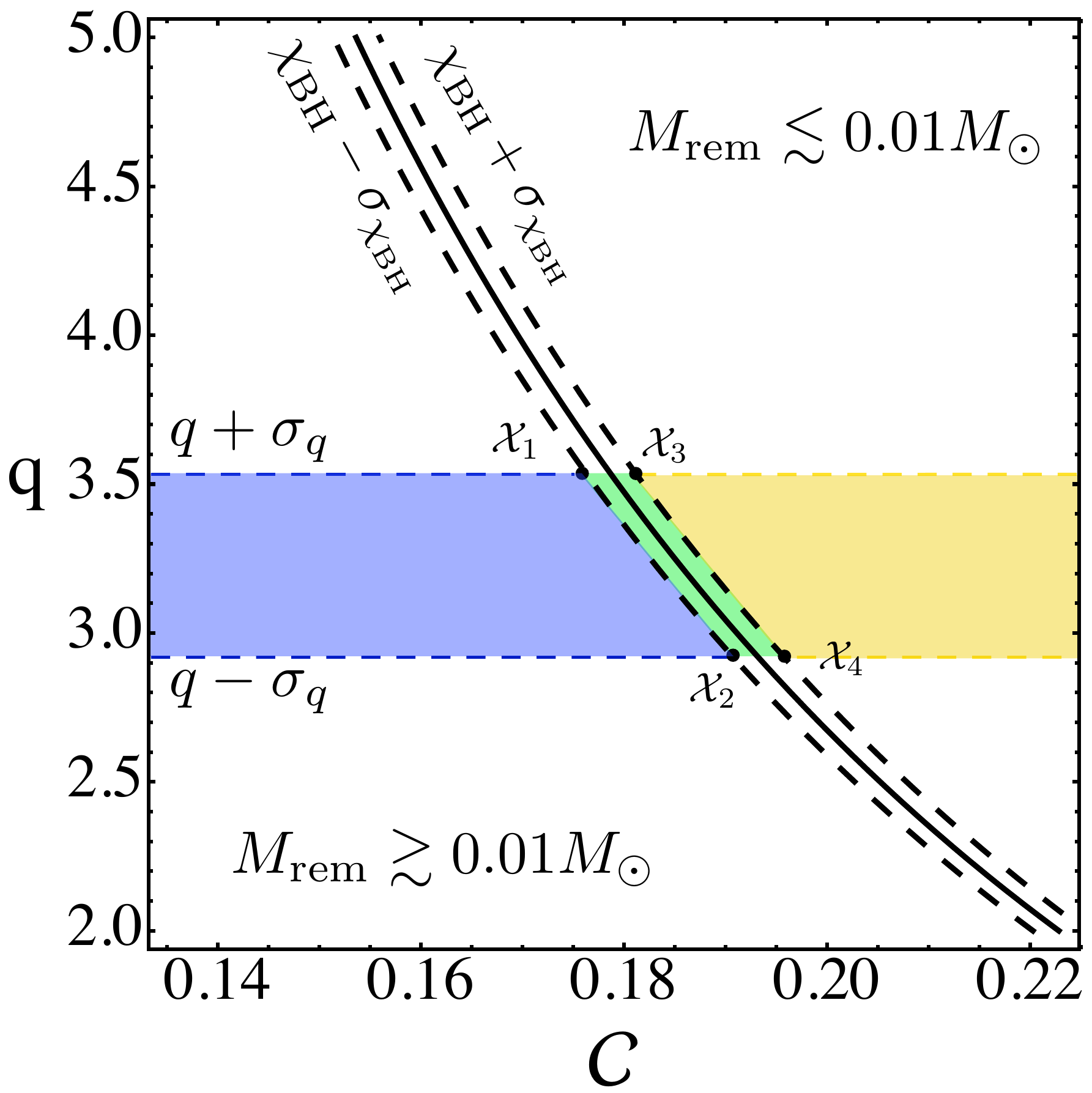}
\caption{Contour plot of the fit (\ref{torusFit}) in the $q$-${\cal C}$
plane, for $M_{\tn{NS}}=1.2M_{\odot}$, $\chi_{\tn{BH}}=0.5$ and
$M_{\tn{rem}} =0.01 M_{\odot}$. The fit  separates the  region allowed
for SGRB ignition (below the fit curve) from the forbidden region (above
the fit). Given the measured values of $q\pm \sigma_q$ and
$\chi_{\tn{BH}}\pm\sigma_{\chi_{\tn{BH}}}$,
a detected signal can correspond to a NS with compactness 
${\cal C}$ which falls  in the blue, green or yellow
region. Since ${\cal C}$ also comes with an error $\sigma_{\cal C}$,
in order to infer if  it can be associated with a SGRB, we need to
evaluate the probability $\tn{P}({\cal C} \leq {\cal
C}_{4})$ and $\tn{P}({\cal C} \leq {\cal C}_{1})$ (see text).
}
\label{FIG1}
\end{figure}

Let us now assume that the gravitational wave signal emitted in a BH-NS
coalescence is detected; a suitable  data analysis  allows to find the
values of the symmetric mass-ratio
$\nu=(M_{\tn{NS}}M_{\tn{BH}})/(M_{\tn{NS}}+M_{\tn{BH}})^{2}$
and of the chirp mass ${\cal M} = \nu^{3/5}(M_{\tn{NS}}+M_{\tn{BH}})$,
from which the mass ratio $q$ can be derived, and of the
black hole spin $\chi_{\tn{BH}}$, with the corresponding errors.
Knowing $q\pm \sigma_{q}$ and
$\chi_{\tn{BH}}\pm\sigma_{\chi_{\tn{BH}}}$, using the fit
(\ref{torusFit}) we can trace the plot of Fig.~\ref{FIG1} in the
$q-{\cal C}$ plane, for an assigned disk mass threshold, say
$M_{\tn{rem}} =0.01 M_{\odot}$.  This plot allows to identify the
parameter region where a SGRB may occur, i.e. the region
$M_{\tn{rem}}\gtrsim 0.01 M_{\odot}$ (below the fit curve in the figure), 
and the forbidden region above the fit ($M_{\tn{rem}}\lesssim 0.01 M_{\odot}$).  
In addition, we
identify four points ${\cal X}_{1},\ldots{\cal X}_{4}$, which are the
intersection between the contour lines for
$\chi_{\tn{BH}}\pm\sigma_{\chi_{\tn{BH}}}$ and the horizontal lines $q\pm \sigma_{q}$.
Let us indicate as ${\cal C}_{1}, \ldots,{\cal C}_{4}$ the corresponding
values of the neutron star compactness. 
Since  the  fit (\ref{torusFit}) is monotonically decreasing, ${\cal
C}_{1}<{\cal C}_{2}<{\cal C}_{3}<{\cal C}_{4}$.  
At this stage we still cannot say whether the detected binary falls in the 
region allowed for the formation of a SGRB or not.

In order to get this information, we need to evaluate 
${\cal C}$. As discussed in \cite{DNV12,DABV13,MGF13,RMSUCF09,Retal13,PROR11}, 
Advanced LIGO/Virgo are expected to measure the gravitational wave phase with 
an accuracy sufficient to estimate the NS tidal deformability $\lambda_2$. Thus, using 
the fit (\ref{Clambda}),  the NS compactness ${\cal C}_{\lambda}$ 
and the corresponding uncertainty 
\begin{equation}
\sigma_{{\cal C}_{\lambda}}^{2}=\sigma^{2}_{\tn{fit}}
+\sum_{i,j}
\frac{\partial {\cal C}_{\lambda}}{\partial p_{i}}
\frac{\partial {\cal C}_{\lambda}}{\partial p_{j}}\tn{Cov}(p_{i},p_{j})
\label{properr}
\end{equation}
can be inferred. In (\ref{properr})
 $p_{i} =\{\lambda_{2},M_{\tn{NS}}\}$, and $\tn{Cov}(p_{i},p_{j})$ is
their covariance. As shown in \cite{MCFGP13},
$\sigma_{\tn{fit}}=0.03\,{\cal C}$  
is the largest relative discrepancy between the value of ${\cal C}$ obtained from the 
fit and the value computed solving the equations of stellar perturbations, for a 
set of EoS covering a large range of stiffness.

Knowing the parameters and their uncertainties, 
the probability that a SGRB is associated to the detected 
coalescence can now be evaluated. 

We assume that $(q,\cal{C}_{\lambda},\chi_\tn{{BH}})$ are described 
by a multivariate Gaussian distribution:
\begin{equation}\label{probab}
{\cal P}(q,{\cal C}_{\lambda},\chi_{\tn{BH}})=\frac{1}{(2\pi)^{3/2}
\vert\Sigma\vert^{1/2}}
\tn{exp}\left[-\frac{1}{2}\Delta^{\tn{T}}\Sigma^{-1}\Delta\right]\ ,
\end{equation}
where $\Delta=(\vec{x}-\vec{\mu})$, ~ $\vec{\mu}=(q,{\cal
C}_{\lambda},\chi_{\tn{BH}})$, ~ $\Sigma$
is the covariance matrix.
Then, we define the maximum and minimum probability 
that the binary coalescence produces an accretion disk with mass 
over the threshold,  $\bar{M}_{\tn{rem}}$, as
\begin{eqnarray}\label{Pminmax}
\tn{P}_{\tn{MAX}}(M_{\tn{rem}}\gtrsim \bar{M}_{\tn{rem}} )
&\equiv&\tn{P}({\cal C}_{\lambda} \leq {\cal C}_{4})\ ,\\ \nonumber
\tn{P}_{\tn{MIN}}(M_{\tn{rem}}\gtrsim \bar{M}_{\tn{rem}} )
&\equiv&\tn{P}({\cal C}_{\lambda} \leq {\cal C}_{1})\ ,
\end{eqnarray}
where $\tn{P}({\cal C}_{\lambda} \leq {\cal C}_{i})$
is the cumulative distribution  
of Eq. (\ref{probab}), which gives  the probability that the measured
compactness ${\cal C}_{\lambda}$, 
estimated through the fit (\ref{Clambda}), is smaller than an 
assigned value ${\cal C}_{i}$.

As an illustrative example,
we now evaluate the probability that a given BH-NS coalescing binary
produces a SGRB, assuming a set of equations of state for the NS matter
and evaluating the uncertainties on the relevant parameters using a
Fisher matrix approach.

{\em  Evaluation of the uncertainties on the binary parameters.}---The
accuracy with which future interferometers will measure a set of binary
parameters $\boldsymbol\theta$, is estimated by comparing the
gravity-wave data-stream with a set of theoretical templates.
For strong enough signals, $\boldsymbol\theta$ are expected to  have a
Gaussian distribution centered around the {\it true values}, with
covariance matrix 
\begin{equation}
\tn{Cov}^{ab}=(\Gamma^{-1})^{ab}\qquad \ ,\qquad \Gamma ^{ab}=\left(
\frac{\partial h}{\partial \theta^{a}}\Bigg\vert\frac{\partial
h}{\partial \theta^{b}}\right)\ .  
\end{equation} 
being $\Gamma$ the Fisher information matrix \cite{PW95}.  $(\cdot
\vert\cdot)$ is the inner product between two GW
templates $\tilde{h}(f)$ and $\tilde{g}(f)$: 
\begin{equation} (h\vert
g)=2\int_{f_{\tn{min}}}^{f_{\tn{max}}}\frac{\tilde{h}(f)\tilde{g}^{\star}(f)
+\tilde{h}^{\star}(f)\tilde{g}(f)}{S_{n}(f)}df\ , 
\end{equation} 
$^{\star}$ denotes complex-conjugation, and $S_{n}(f)$ is the noise
spectral density of the considered detector. To model the waveform we
use the TaylorF2 approximant in the frequency domain, assuming the
stationary phase approximation \cite{DIS00}: 
\begin{equation} 
h(f)={\cal
A}(f)e^{i\psi(f)}=\sqrt{\frac{5}{24}}\frac{{\cal
M}^{5/6}}{\pi^{2/3}d}f^{-7/6}e^{i\psi(f)}\ .  
\end{equation} 
where $d$ is the source distance.  The post-Newtonian expansion of the
phase includes spin-orbit and tidal corrections.  It can be
written as 
$\psi(f)=\psi_{\tn{PP}}+\psi_{\tn{T}}$, where the point-particle term is
\begin{equation}\label{PPphase} 
\psi_{\tn{PP}}(f)=2\pi
t_{c}-\phi_{c}-\frac{\pi}{4}+\frac{3}{128}({\cal M}\pi f)^{-5/3}
\sum_{i=0}^{7}e_{i}(m\pi f)^{i/3}\ , 
\end{equation} 
and  $t_{c}$ and
$\phi_{c}$ are the time and the phase at coalescence.  The coefficient
$e_{i}$ are listed in \cite{BFIJ02,BDE04}. The tidal contribution
$\psi_{\tn{T}}$
 is given by \cite{VFH11,DNV12}
\begin{align}\label{Tphase}
\psi_{\textnormal{T}}(f)=-&\frac{117\Lambda}{8 \nu m^5}
x^{5/2}\Bigg[1+\frac{3115}{1248}x-\pi x^{3/2}\ +\nonumber\\
&+\left(\frac{23073805}{3302208} +\frac{20}{81}6\right)x^{2}
-\frac{4283}{1092}\pi x^{5/2}\Bigg]\ , 
\end{align} 
where $x=(m \pi f)^{5/3}$ and $\Lambda$ is the averaged tidal
deformability, which for BH-NS binaries reads \cite{FH08}:
$\label{lambda} \Lambda=\lambda_{2}
\frac{\left(1+12q\right)}{26}$.

We consider non rotating
NSs, as this is believed to be a reliable approximation of
real astrophysical systems \cite{BC92,K92}.  The GW waveform is
therefore described in terms of the following set of parameters,
$\boldsymbol\theta=(\ln{\cal
A},t_{c},\phi_{c},\ln{\cal M},\ln\nu,\Lambda,\beta)$,
where $\beta$ is the 2 PN spin-orbit contribution in
$\psi_{\tn{PP}}$: 
\begin{equation}
\beta=\frac{\chi_{\tn{BH}}}{12}\left[113\frac{M_{\tn{BH}}^2}{m^2}+75\nu\right]
\hat{\textbf{L}}\cdot\hat{\textbf{S}}_{\tn {BH}}\ , 
\end{equation} 
where $\hat{\textbf{L}}$, $\hat{ \textbf{S}}_{\tn{NH}}$ are the unit
vectors in the direction of the orbital angular momentum and of the
spin, respectively. We choose the BH spin  aligned with
the orbital angular momentum.  Moreover, since ${\chi}_{\tn {BH}}\leq
\vert1\vert$, $\beta \lesssim 9.4$; therefore we consider the
following prior probability distribution on $\beta$:
$p^{(0)}(\beta)\propto \tn{exp}\left[-\frac{1}{2}\left(\beta/
9.4\right)^{2}\right]\ .$

Thus, we need to  compute a $7\times 7$ Fisher matrix. 
However, since $\ln {\cal A}$ is uncorrelated 
with the other variables, we perform derivatives only with 
respect to the remaining six parameters 
$\boldsymbol\theta=(t_{c},\phi_{c},\ln{\cal M},\ln\nu,\Lambda,\beta)$.
In our analysis we consider both second (AdvLIGO/Virgo) and 
third generation (Einstein Telescope, ET, \cite{ET1})
detectors. For AdvLIGO/Virgo we use the \texttt{ZERO\_DET\_high\_P} 
noise spectral density of AdvLIGO \cite{zerodet}, in the frequency
ranges $[20\ \textnormal{Hz},f_{\textnormal{ISCO}}]$;
for the Einstein Telescope we use  the analytic fit of the  sensitivity
curve provided in \cite{ET2}, in the range 
 $[10\ \textnormal{Hz},f_{\textnormal{ISCO}}]$.
$f_{\tn{ISCO}}$ is the frequency of the Kerr ISCO including
corrections due to NS self-force \cite{Fav11}:
\begin{equation}\label{fisco}
f_{\tn{ISCO}}=\frac{M_{\tn{BH}}}{m\pi }\Omega(\chi_{\tn{BH}})
\left[1+\nu c^{\tn{GSF}}(\chi_{\tn{BH}})\right]\ ,
\end{equation}
with $\Omega(\chi_{\tn{BH}})=\tn{sign}(\chi_{\tn{BH}})M_{\tn{BH}}^{1/2}/
(r_{\tn{ISCO}}^{3/2}+\chi_{\tn{BH}})$.

We model the NS structure by means of piecewise polytropes, 
 \cite{RMSUCF09}. 
The core EoS changes with an overall pressure shift $p_{1}$, specified at the density 
$\rho_{1}=5.0119\times 10^{14}$g cm$^{-3}$. Once the adiabatic index $
\Gamma_{\textnormal{core}}$ is fixed, increasing $p_{1}$ produces a family of 
neutron stars with growing radius for a given mass. Choosing 
$\Gamma_{\textnormal{core}}=3$ and $p_1 = (10^{13.95},10^{13.55},10^{13.45},
10^{13.35})$ g/cm$^3$ we obtain four EoS, \texttt{2H},\texttt{H},
\texttt{HB} and \texttt{B}, which denote very stiff, stiff, moderately stiff and soft 
EoS, respectively.  The  stellar parameters  for 
$M_{\tn{NS}}=(1.2,1.35)M_{\odot}$,
 are shown in Table~\ref{Tableconf}.

\begin{table}[h!]
\centering
\begin{tabular}{c|ccc||cccc}
\hline
\hline
\texttt{EoS} & $M_{\tn{NS}} (M_{\odot})$ & ${\cal C}$  
& $\lambda_{\tn{2}}$ (km$^5$) & $M_{\tn{NS}} (M_{\odot})$ & ${\cal C}$  
& $\lambda_{\tn{2}}$ (km$^5$)\\
\hline
 \texttt{2H} & 1.2  & 0.117  & 75991 & 1.35  & 0.131  & 72536 \\
 \texttt{H}   & 1.2  & 0.145  & 21232 & 1.35  & 0.163  & 18964\\
 \texttt{HB} & 1.2  & 0.153  & 15090 & 1.35  & 0.172  & 13161\\
 \texttt{B}   & 1.2  & 0.162  & 10627 & 1.35  & 0.182  & 8974\\
 \hline
 \hline
\end{tabular}
\caption{For each EoS we show the NS mass, the compactness 
${\cal C}=M_{\tn{NS}}/R_{\tn{NS}}$, and the tidal 
deformability $\lambda_{2}$.}
\label{Tableconf}
\end{table}


{\em Numerical results.}---Following the strategy previously outlined,
we compute the minimum and maximum probabilities (\ref{Pminmax}) 
that the coalescence of a BH-NS system produces a 
remnant disk with mass above a threshold  $\bar{M}_{\tn{rem}}$, 
for the NS models listed in Table~\ref{Tableconf} 
and different values of the mass ratio $q$. The results are given in 
Table~\ref{table2}, for  
$q=3$ and $q=7$, black hole spin $\chi_{\tn{BH}}=(0.2,0.5,0.9)$,
$M_\tn{NS}=(1.2,1.35)~M_\odot$, and  disk mass
thresholds $\bar{M}_{\tn{rem}}=0.01M_{\odot}$.

For AdvLIGO/Virgo we put the source at a distance of $100$ Mpc.
For ET the binary is at $1$ Gpc. In this case the signal must be
suitably redshifted \cite{CF94,MGF13}, and  we have assumed that $z$ is known
with a fiducial error of the order of 10\% \cite{MR12}.


\begin{table*}[ht]
\centering
\begin{tabular}{cc|ccc|ccc||ccc|ccc}
& & $q=3$ & $d=100$ Mpc & AdV & $q=3 $ & $d=1$ Gpc & ET 
& $q=7 $ & $d=100$ Mpc & AdV & $q=7 $ & $d=1$ Gpc & ET\\
\hline
& $M_{\tn{NS}}=1.2M_{\odot}$&
& $\chi_{\tn{BH}}$ &   & & $\chi_{\tn{BH}}$ &&& $\chi_{\tn{BH}}$ &&& $\chi_{\tn{BH}}$\\
\hline
\texttt{EOS} & ${\cal C}_{\lambda}$ & $0.2 $ 
& $0.5$ & $0.9$ & $0.2 $ & $0.5$ & $0.9$ & $0.2 $ & $0.5$ & $0.9$ & $0.2 $ & $0.5$ & $0.9$\\
\hline
\texttt{2H} & 0.118 & 1 & 1 & 1 & 1 & 1 & 1 & 0.4 & [0.8-0.9]  &  1 & [0.3-0.4] & [0.7-0.8] & 1 \\
\texttt{H}  &  0.147 & [0.6-0.9] & 1 & 1 & [0.9-1] & 1 & 1 & 0.4 & 0.4  & [0.8-0.9] & 0.3 & [0.3-0.4] & 1\\
\texttt{HB} & 0.155 & [0.5-0.7] & [0.9-1] & 1 &  [0.7-0.8] & 1 & 1 & 0.4 & 0.4 & [0.7-0.8]  & 0.3 & 0.3 & 0.9 \\
\texttt{B}  &  0.164 & [0.4-0.6] & [0.7-0.8] & 1 & [0.5-0.6] & 1 & 1 & 0.4 & 0.4 & [0.6-0.7] & 0.3 & 0.3 & [0.7,0.8] \\
\hline
& $M_{\tn{NS}}=1.35M_{\odot}$ &&&&&&&&&&\\
\hline
\texttt{2H} & 0.132 & 1 & 1 & 1 & 1 &1 &1 & 0.3 & [0.4-0.5] & 1 & 0.2 & [0.4,0.5] &1 \\
\texttt{H}  & 0.164 & [0.4-0.6] & [0.8-0.9] & 1 & [0.4-0.6] & 1 &1 & 0.4 & 0.4  & 0.7 & 0.3 & 0.3 & 0.8 \\
\texttt{HB} & 0.173 & [0.4-0.5] & [0.6-0.8] & 1 & [0.3-0.4] & 0.8 &1 & 0.4 & 0.4 & 0.6  & [0.3-0.4] & [0.3-0.4] & 0.6 \\
\texttt{B}  & 0.184   & [0.4-0.5] & [0.5-0.6] & 1 & [0.3-0.4] & 0.6 &1 &  0.4  & 0.4 & 0.5 & 0.4 & 0.4 & [0.5-0.6] \\
\end{tabular}
\caption{We show the probability range [P$_{\tn{MIN}}$,P$_{\tn{MAX}}$]
that the coalescence of a BH-NS binary produces a disk mass
larger than $\bar{M}_{\tn{rem}}=0.01M_{\odot}$ for AdLIGO/Virgo (AdV) 
and for the Einstein telescope (ET), for binaries with $q=3$ and $q=7$, 
NS masses (1.2,1.35)$M_{\odot}$, and BH spin $\chi_{\tn{BH}}=(0.2,0.5,0.9)$.
Sources are assumed to be at $d=100$ Mpc for advanced detectors, and 
$d=1$ Gpc for the Einstein Telescope. The star compactness
${\cal C}_{\lambda}$ is estimated throughout the universal relation
(\ref{Clambda}).}\label{tableres1}
\label{table2}
\end{table*}

The first clear result is that as the BH spin
approaches the highest value 
we consider, $\chi_{\tn{BH}}=0.9$,  and for low mass ratio $q=3$,
the probability that a BH-NS coalescence produces a disk with
mass above the threshold  is insensitive to the NS
internal composition, and it approaches  unity for all 
considered configurations. These would be good candidates for
GRB production.
For the highest mass ratio we consider,  $q=7$, the probability to form a sufficiently 
massive disk depends on the NS mass and EoS, and on the detector.
In particular, it decreases as the EoS softens, and as the NS mass
increases. This is a general trend, observed also for smaller values of
$\chi_{\tn{BH}}$.   
However, when $\chi_{\tn{BH}}=0.9$ the probability that the coalescence
is associated to a SGRB is always $\gtrsim 50\%$ .

Let us now consider  the results for $\chi_{\tn{BH}}=0.2$.  
If the NS mass is $1.2~M_\odot$ the probability that a detected GW
signal from a BH-NS coalescence is associated to the formation of a
black hole with a disk of mass above threshold is $\gtrsim 50\%$ for
both AdvLIGO/Virgo and ET, provided $q=3$. For larger  NS mass,
this remains true only if the NS equation of state is stiff (\texttt{2H}
or \texttt{H}).
High values of $q$ are disfavoured.

When the  black hole spin has an intermediate value, say
$\chi_{\tn{BH}}=0.5$,  Table~\ref{table2} shows that, 
the NS compactness plays 
a key role in the identification of good candidates for GRB production,
for both detectors. Again large values of the mass ratio yield small
probabilities.

The range of compactness shown in  Table~\ref{table2}
includes neutron stars with radius ranging within $\sim [10,15]$ km.
From the table it is also clear that if we choose a
compactness smaller than the minimum value,
the probability of generating a SGBR increases, and
the inverse is true if we consider compactness larger than our maximum.


{\em Concluding remarks.}---

The method developed in this paper can be used in several different ways.
In the future, gravitational wave detectors are expected to reach a
sensitivity sufficient to extract the parameters on which our analysis
is based, i.e. chirp mass, mass ratio, source distance, spin and tidal
deformability. We can also expect  that the steady improvement of the 
efficiency of computational facilities experienced in recent years will
continue, reducing the time needed to obtain these parameters from a
detected signal. Moreover, the higher sensitivity  
will allow to detect sources in
a much larger volume space, thus increasing the detection rates.
In this perspective, the method we envisage in this paper will be useful
to trigger the electromagnetic follow-up of a GW detection, 
searching for the afterglow  emission of a SGRBs.

Waiting for the future, the method we propose can be used 
in the data analysis of advanced detectors as follows:
\begin{itemize}
\item
Table II indicates the systems which are more likely to
produce accretion disks sufficiently massive to generate a SGRB.
The table can be enriched including more 
NS equations of state or more binary parameters;
however, it already contains a clear information on which is
the range of  parameters to be used in the GW data analysis,
if the goal is  to  search for  BH-NS signals which may be associated to a
GRB. For instance, 
Table II suggests that searching for mass ratio smaller than, or equal
to, 3-4, and values of the black hole angular momentum larger than
0.5-0.6 would allow to save time and computational resources in low
latency search. In addition, it would allow to gain sensitivity in externally
triggered searches performed in time coincidence with short GRBs
observed by gamma-ray satellites.
\item
If a SGRB is observed sufficiently close to us in the electromagnetic
waveband, the parameters of the GW signal detected in coincidence would
allow to set a threshold on the mass of the accretion disk.  If the GW
signal comes, say, from a system with a BH with spin
$\chi_{\tn{BH}}=0.5$, mass ratio $q=7$, and neutron star mass
$M_{\tn{NS}}=1.2 M_\odot$, from Table~\ref{table2},  equations of state
softer than the EoS \texttt{2H} would be disfavoured.  Thus, we would be
able to shed light on the dynamics of the binary system, on its
parameters and on the internal structure of its components. We would
enter into the realm of gravitational wave astronomy.  
\end{itemize}
Finally,  it is worth stressing that as soon as the fit 
(\ref{torusFit}) will be extended to NS-NS coalescing
binaries, this information will be easily implemented in
our  approach.
Being the rate of NS-NS coalescence higher than that of BH-NS, our
approach will acquire more significance, and will
be a very useful tool to study these systems.


{\em Acknowledgements.}---We would like to thank M. Branchesi for 
her comments on the paper, which helped us to  clarify several issues.
We would like to thank L.~Pagano, 
R.~Amato and M.~Muccino for useful discussions and comments. 
A.~M. is supported by a "Virgo EGO Scientific Forum" (VESF) grant. 
Partial support comes from the CompStar network, COST Action MP1304.



\bibliographystyle{h-physrev4}

\end{document}